# Measuring direct flexoelectricity at the nanoscale

**Authors:** Daniel Moreno-Garcia[1], Luis Guillermo Villanueva[1]

[1] Advanced NEMS Group, École Polytechnique Fédérale de Lausanne (EPFL), Lausanne 1015, Switzerland

## 1.1 Abstract

Flexoelectricity is a property of all dielectric materials, where inhomogeneous strain induces electrical polarization. This effect becomes particularly prominent at the nanoscale where larger strain gradients can be obtained. While flexoelectric charges have been measured in mm-scale systems, direct measurements in nanoscale-thickness materials have not yet been achieved. Given that one of the most prominent applications of flexoelectricity is in nano-electro-mechanical systems (NEMS), confirming the presence and magnitude of the effect at these scales is essential. This study presents the first-ever measurements of flexoelectric-generated charges (direct effect) in nanoscale-thickness materials, using cantilevers with a 50 nm hafnium oxide layer. We confirm that the estimated flexoelectric coefficient from said measurements aligns with the values obtained from complementary experiments using the flexoelectric inverse effect. Additionally, by changing the cantilever geometry (modifying the width of the cantilevers), we demonstrate a 40% increase in the effective flexoelectric coefficient, explained by the interplay of different flexoelectric tensor components. These findings not only validate the presence of flexoelectric effects at the nanoscale but also open the possibility for full flexoelectric transduction of the motion in NEMS/MEMS devices.

## 1.2 Introduction

Flexoelectricity is a property of dielectric materials where an inhomogeneous mechanical deformation, such as bending, induces electrical polarization (direct effect) [1]. Conversely, an applied electric field induces mechanical deformation (inverse effect) [2]. Unlike piezoelectricity, which requires non-centrosymmetric materials, flexoelectricity can occur in all dielectric materials [3]. This effect is particularly significant at the nanoscale, where large strain gradients are easier to achieve [4]. The simplest expression of this relationship is captured in Equation 1.1, where the flexoelectric coefficient is represented by $\mu$, the material strain gradient is $\partial \varepsilon / \partial x$ and the generated polarization is $P$ [5].

$$P = \mu \frac{\partial \varepsilon}{\partial x} \qquad (1.1)$$

A conceptual visualization of flexoelectricity can be illustrated using a simple ionic lattice (Figure 1.1a). When the material bends, the strain gradient causes a relative displacement between positively and negatively charged ions, leading to a local redistribution of charges and inducing a net polarization (Figure 1.1c). In materials where ionic movement is minimal, flexoelectricity can also be understood as a redistribution of electron density under strain gradients [6], enabling the effect in any type of dielectrics.

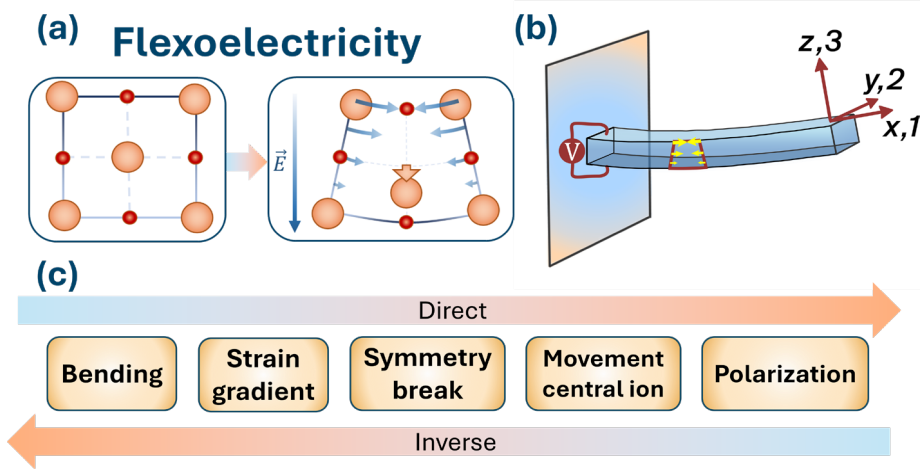

**Figure 1.1 Schematic representation of flexoelectricity.** a) An ionic lattice structure before (left) and after deformation (right), where the strain gradient induces polarization. b) In a cantilever, bending generates a voltage across the beam's thickness. c) Diagram of the steps by which an ionic lattice produces polarization from bending (direct effect) or bending in response to an applied voltage (inverse effect).

Direct flexoelectricity in a mechanical structure (Figure 1.1b) generates surface charges across the beam when it bends along its length. To facilitate the reading of small charges, studies in the literature tend to use samples of large lateral dimensions (millimeters to centimeters). Consequently, to avoid breaking the sample when applying the load, the samples have thicknesses that range from 0.5 to 3 mm.

Typically these measurements employ a three- or four-point bending tester, connected to an amplifier [7], [8], [9], [10], [11]. Other authors produce the strain in the material with a loudspeaker and measure charges with a lock-in amplifier. In these cases, the displacement of the loudspeaker is measured using a Displacement Voltage Ratio Transformer (DVRT) [1], [12], [13], [14] or an optical sensor [15], [16]. Additionally, some approaches utilize a piezoelectric shaker to flex the sample, with the displacement controlled by a Laser Doppler Vibrometer (LDV). In these cases, charge measurements are conducted using an amplifier [17], [18].

To date, no method reported in the literature has successfully measured direct flexoelectric effect in materials with nanoscale thicknesses. This is primarily due to the challenges in achieving sufficient sensitivity and accurately controlling strain gradients at such small scales [2]. Given that one of the most promising applications of flexoelectricity lies in Nano-Electro-Mechanical Systems (NEMS), where device dimensions are on the nanometer scale, confirming the presence and magnitude of the flexoelectric effect at these scales is crucial. This could enable the development of novel sensors, actuators, and energy-harvesting devices that use nanoscale flexoelectricity [19].

## 1.3 Methodology

Our approach for measuring flexoelectric charges at the nanoscale starts by fabricating cantilevers with a thin (50 nm) layer of hafnium oxide, sandwiched between two platinum electrodes (20 nm). Hafnium oxide is selected due to its high dielectric constant (within microelectronic compatible materials), well-characterized coefficient [20], [21], and its wide availability in cleanroom facilities. The cantilevers measure 20 µm in length and vary in width from 5 µm to 50 µm (Figure 1.2). Even though we favor narrower devices, we include a range of widths in this study to facilitate charge detection.

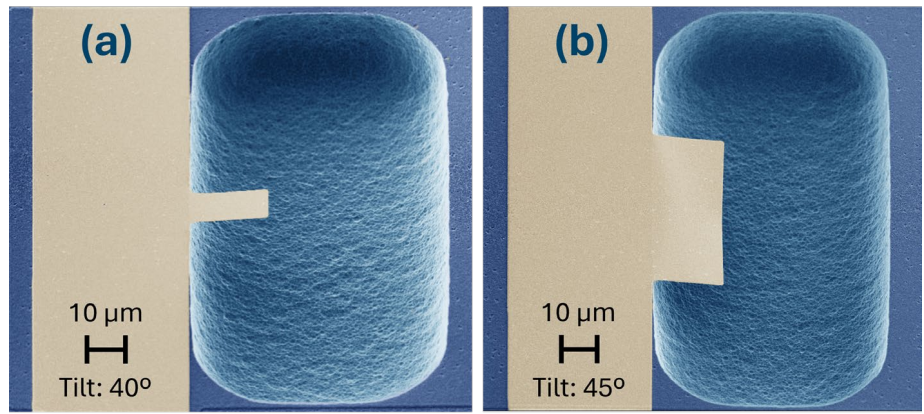

**Figure 1.2 SEM images of fabricated cantilevers.** a) Cantilever with a length of 20 µm and a width of 5 µm. b) Cantilever with a length of 20 µm and a width of 50 µm.

To maximize the flexoelectric response, we excite the cantilevers at their resonance frequency using a piezoelectric shaker under the cantilevers chip. Operating at resonance amplifies the mechanical displacement, improving the signal-to-noise ratio. Measurements are conducted under vacuum ($3 \cdot 10^{-4}\ mbar$) to reduce air damping, further enhancing displacement amplitude. The generated flexoelectric charges are collected by platinum electrodes and amplified using a low-noise voltage amplifier (Sierra Amps [22]) placed in close proximity to minimize the load capacitance and signal loss. The experimental setup (Figure 1.3) includes a Laser Doppler Vibrometer to monitor cantilever displacement. All signals are processed through a lock-in amplifier (Zurich Instruments UHFLI), which controls the shaker, reads the electrical charges, and monitors the cantilever's displacement. Importantly, the electrical charges and the vibrometer signal are not monitored simultaneously, but subsequently.

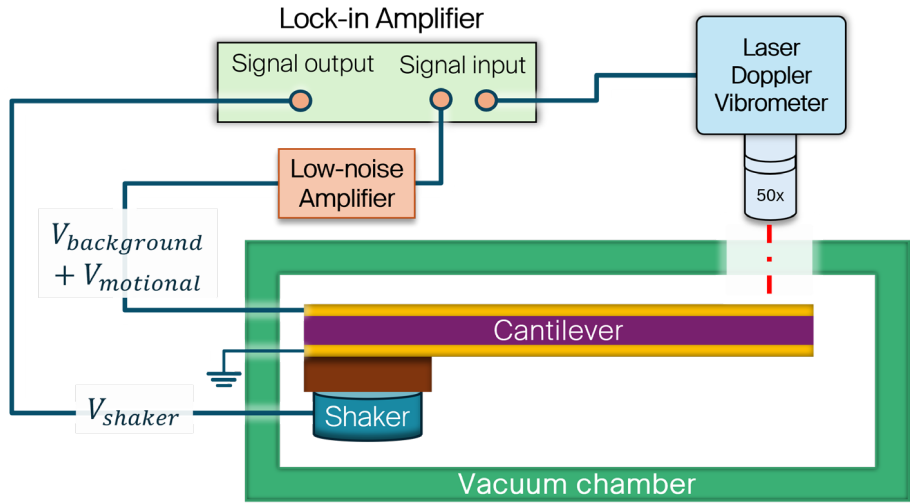

**Figure 1.3 Schematic of the experimental setup to measure flexoelectric charges in nanoscale-thickness materials.** The cantilever is actuated by a piezoelectric shaker at its resonance frequency under vacuum. A low-noise amplifier amplifies the generated flexoelectric charges ($V_{motional}$) and parasitic charges ($V_{background}$). A Laser Doppler Vibrometer monitors the displacement of the cantilever. All signals are processed in a lock-in amplifier.

The first measurement quantifies the cantilever's displacement for a certain shaker actuation. By sweeping the excitation frequency around the cantilever's resonance frequency, we measure the displacement at the laser's measurement point (Figure 1.4). However, our primary interest is the displacement at the cantilever's tip, which is essential for calculating the strain gradient (beam curvature). To extrapolate the tip displacement from the measured point, we analyze the thermomechanical noise and use a standard cantilever beam model. This method allows us to translate the measurements at any point along the cantilever to the tip [23], [24].

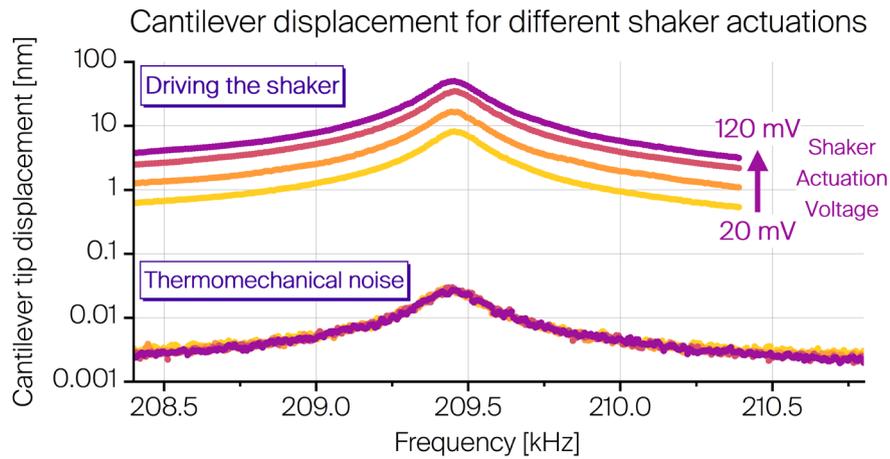

**Figure 1.4 Laser Doppler Vibrometer measurements of cantilever displacement around the resonance frequency.** The displacement amplitude is linearly proportional to the actuation voltage applied to the shaker. Thermomechanical noise measurements, conducted without actuation, are used to verify the laser position on the cantilever and translate the amplitude measurements to cantilever's tip displacement.

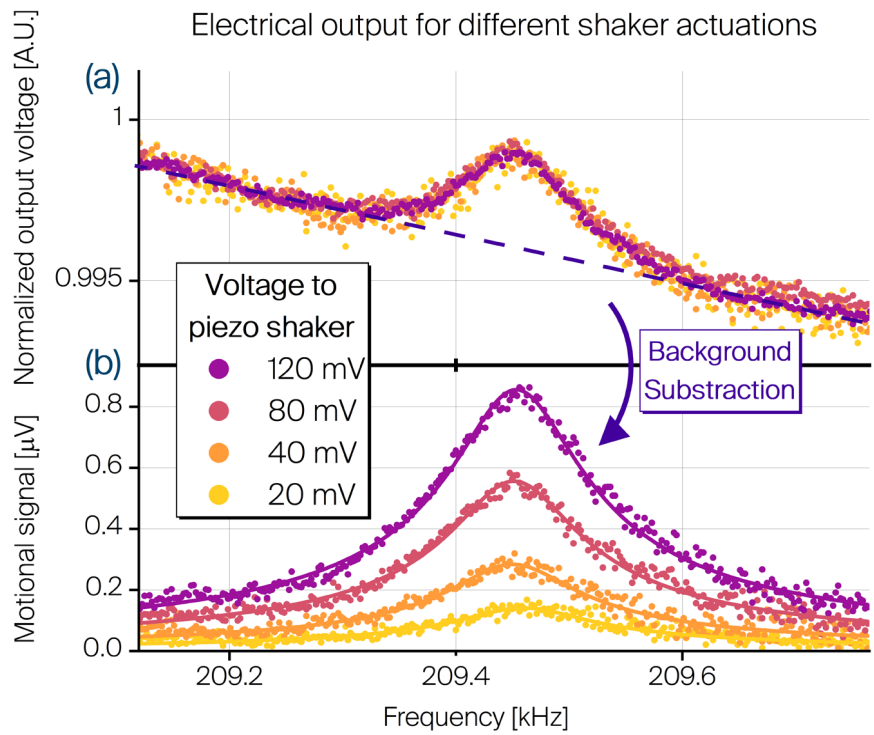

**Figure 1.5 Flexoelectric voltage generated in the cantilever around its resonance frequency.** a) Normalized output voltage for a cantilever with a width of 40 µm and a length of 20 µm. We observe the flexoelectric charges superimposed on the background signal. b) Lorentzian signal corresponding to the cantilever's motion, obtained by subtracting the background signal from the output signal.

The second measurement evaluates the electrical charges generated by the flexoelectric effect. We perform the measurements without the laser, as its illumination can induce photoelectric effects and interfere with the flexoelectric charges [25]. Figure 1.5 shows the electrical measurements obtained from a cantilever with dimensions of 40 μm in width and 20 μm in length. In Figure 1.5a, the normalized transfer function is plotted as a function of frequency, revealing the cantilever's resonance peak superimposed on a background signal. This background arises from parasitic capacitance pathways between the shaker's actuation and the cantilever. To isolate the flexoelectric signal, we perform a linear baseline subtraction for both components in phase and in quadrature with the drive, to remove the background contribution, resulting in the cleaned data presented in Figure 1.5b. The extracted signal corresponds to the flexoelectric charges generated due to the cantilever's motion, which are linearly dependent on the shaker's actuation.

By combining the measured electrical amplitudes ($V_{motional}$) with the cantilever displacements at the tip ($u$) at various shaker actuations, we calculate the effective flexoelectric coefficient ($\mu_{eff}$) using Equation 1.2. This calculation also involves the cantilever's width ($W$), the capacitance of the measurement path leading to the low-noise amplifier ($C_{load}$) and the derivative of the normalized n-mode shape ($\phi'_n(L)$). For the cantilever's first mode, $\phi'_1(L) = -1.3765/L$, where $L$ represents the length of the cantilever. Detailed derivation is provided in the supplementary material.

$$\mu_{eff} = \frac{C_{load} \cdot V_{motional}}{W \cdot \phi'_n(L) \cdot u} \tag{1.2}$$

## 1.4  Results and discussion

Figure 1.6 presents the effective flexoelectric coefficient for cantilevers of varying widths. We observe that $\mu_{eff}$ increases with increasing width, reaching a saturation value for widths larger than 40 μm. This trend is attributed to a regime shift, as the cantilevers transition from the $W < L$ condition to $W > L$. This change affects how the flexoelectric tensor components $\mu_{11}$ and $\mu_{12}$ [4] interact to determine the generated charges. Table 1.1 describes the theoretical asymptotic limits of the effective flexoelectric coefficient for both narrow and wide cantilevers (compared to the length). Detailed derivations are provided in the supplementary material. By making the ratio between the limits we obtain $\mu_{eff,w}/\mu_{eff,n} = 1/(1-\nu)$, which depends solely on the material's Poisson's ratio ($\nu$), and not on the flexoelectric tensor terms. For hafnium oxide, with $\nu = 0.3$, the ratio evaluates to approximately 1.42. This theoretical prediction aligns well with the data in Figure 1.6, where increasing the width results in a 42% increase in $\mu_{eff}$.

| **Narrow Cantilevers** $t \ll W \ll L$ | **Wide Cantilevers** $t \ll L \lesssim W$ |
|---|---|
| $\mu_{eff,n} = -\nu\mu_{11} + \mu_{12}(1-\nu)$ | $\mu_{eff,w} = \mu_{11}\dfrac{\nu}{\nu-1} + \mu_{12}$ |
| $\mu_{eff,w}/\mu_{eff,n} = 1/(1-\nu)$ ||

**Table 1.1 Asymptotical effective flexoelectric coefficients.** Asymptotical limits of the effective flexoelectric coefficient for narrow and wide cantilevers. The ratio between these limits is independent of the flexoelectric tensor components, depending only on Poisson's ratio of the material. (t = Cantilever's thickness).

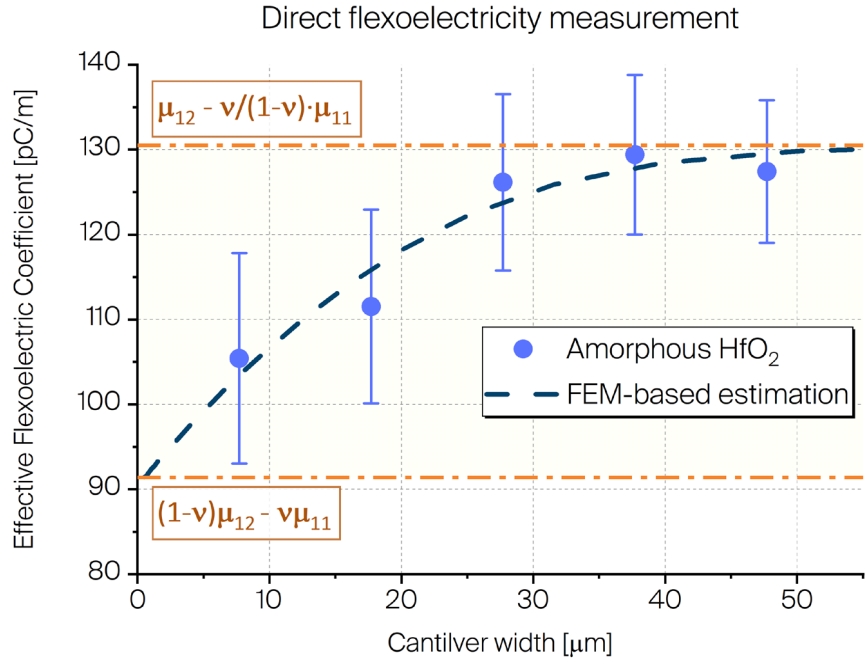

**Figure 1.6 Effective flexoelectric coefficient for cantilevers with different widths.** The growth in the effective flexoelectric coefficient is theoretically expected, as the interplay between the tensorial components depend on the width of the cantilever. The figure shows the asymptotical limits for the extreme cases of the width.

The direct flexoelectric measurements presented in this work (Figure 1.6) align with the published coefficients we previously obtained from inverse flexoelectric measurements [20]. In that study, we report a value of $\mu_{eff} = 105 \pm 10\ pC/m$ for narrow cantilevers, which aligns closely with the direct flexoelectric measurements.

## 1.5 Conclusion

This study presents the first flexoelectric charge measurements in nanometrically thin materials. We employ microcantilevers with a 50 nm layer of hafnium oxide between platinum electrodes. Electrical signals are captured using a low-noise amplifier and they are compared with the cantilever displacement which is measured with a Laser Doppler Vibrometer. The results align with hafnium oxide coefficient values obtained through inverse methods. Additionally, we demonstrate that the effective flexoelectric coefficient increases with cantilever width (for a fixed length), showing the dependence on cantilever geometry. These findings could help in the fabrication of next-generation NEMS devices.

## 1.6 Supplementary Material

### *1.6.1 Flexoelectric formula derivation*

Derivation of the formula to convert the read voltage (after background subtraction) and cantilever tip displacement into effective flexoelectric coefficient.

| $J_D$: current density | P: polarization | $I_{canti}$: cantilever's current |
|---|---|---|
| L : cantilever's length | W: cantilever's width | A = L · W: cantilever's area |
| $\omega = 2\pi f$: frequency | $\mu_{eff}$: flexoelectric coefficient | $\varepsilon_{11}$: strain, along length |
| x: coordinate along cantilever's length | z: coordinate along cantilever's thickness | $R_{ampli}$: amplifier's resistance |
| u: vibrational amplitude at cantilever's tip | $\phi_n$: Normalized n-mode shape | $R_p$: device resistance |
| $V_{motional}$: voltage from flexoelectric charges | Gain: low-noise amplifier's gain | $\beta_n$: wavenumber of mode n |
| $C_{device}$: capacitance cantilever and PCB | $C_{connector}$: capacitance connector | $C_{ampli}$: capacitance amplifier |

$$J_D = \frac{\partial P}{\partial t} = j\omega P \tag{1.3}$$

$$I_{canti} = \int J_D \, dA = \int j\omega P \, dA = j\omega W \int_0^L P \, dx \tag{1.4}$$

$$P_3 = \mu_{eff} \frac{\partial \varepsilon_{11}}{\partial z} = \mu_{eff} \cdot \phi_n''(x) \cdot u \tag{1.5}$$

Combining 1.4 and 1.5:

$$I_{canti} = j\omega \, W \, \mu_{eff} \, u \int_0^L \phi'' \, dx = j\omega \, W \, \mu_{eff} \, u[\phi_n'(L) - \phi_n'(0)] \tag{1.6}$$

The $\phi'(0)$ should be 0, because it is the anchor point of the cantilever.

$$I_{canti} = j\omega \, W \, \mu_{eff} \, u \, \phi_n'(L) \tag{1.7}$$

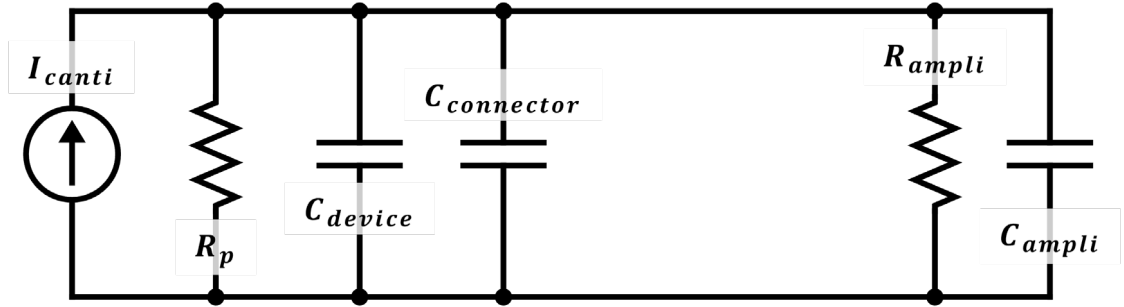

**Figure 1.7 Equivalent circuit of the electronic setup.** The cantilever is represented as a current source ($I_{canti}$), which has an equivalent parallel resistance ($R_p$) and capacitance ($C_{device}$). The capacitance of the connector is represented by ($C_{connector}$), the input low-noise amplifier resistance ($R_{ampli}$) and capacitance ($C_{ampli}$).

$$I_{canti} = \left(\frac{1}{R_p} + \frac{1}{R_{ampli}} + j\omega(C_{device} + C_{connector} + C_{ampli})\right) V_{motional} \qquad (1.8)$$

If we assume: $\frac{1}{R_p} + \frac{1}{R_{ampli}} \ll 1$

$$I_{canti} = j\omega(C_{device} + C_{connector} + C_{ampli}) \cdot V_{motional} \qquad (1.9)$$

$$C_{load} = C_{device} + C_{connector} + C_{ampli} \qquad (1.10)$$

Equalizing 1.7 and 1.9:

$$\mu_{eff} = \frac{C_{load} \cdot V_{motional}}{W \cdot \phi_n'(L) \cdot u} \qquad (1.11)$$

The mode profile function:

$$\phi_n(x) = \frac{1}{2}\left[\cos(\beta_n x) - \cosh(\beta_n x) - \frac{\cos(\beta_n L) + \cosh(\beta_n L)}{\sin(\beta_n L) + \sinh(\beta_n L)}(\sin(\beta_n x) - \sinh(\beta_n x))\right] \qquad (1.12)$$

$$\phi_n'(x) = \frac{1}{2}\left[-\beta_n \cdot \sin(\beta_n x) - \beta_n \cdot \sinh(\beta_n x) - \frac{\cos(\beta_n L) + \cosh(\beta_n L)}{\sin(\beta_n L) + \sinh(\beta_n L)} \cdot \beta_n \cdot (\cos(\beta_n x) - \cosh(\beta_n x))\right] \qquad (1.13)$$

For the first mode of vibration (n=1), $\beta_1 = 1.8751/L$:

$$\phi_1'(L) = -\frac{1.3765}{L} \qquad (1.14)$$

### 1.6.2 *Measurement uncertainty evaluation*

Extended formula for the direct measurement of flexoelectricity, including all dependencies.

$$\mu_{direct} = \frac{(C_{device} + C_{connector} + C_{ampli}) \cdot L}{1.3765 \cdot W} \cdot \frac{V_{motional}}{Amp_{sweep} \cdot \sqrt{\frac{4 \cdot Q \cdot K_B T}{\frac{1}{4} L \cdot W \cdot \Sigma \rho_i t_i \cdot (2\pi f_R)^3 \cdot PSD/2}}} \quad (1.15)$$

A summary of the variables used in our measurements is provided in the following table:

| Variable | Method | Value range | Uncertainty | Contrib. |
|---|---|---|---|---|
| Q | | ~1500 | 1 % | 0.5 % |
| $f_R$ | Output from LDV | 100 kHz − 2 MHz | 0.01% | 0.015 % |
| PSD | | ~$10^{-8}$ V$^2$/Hz | 1 % | 0.5 % |
| $Amp_{sweep}$ | | ~100 mV | 0.1 % | 0.1 % |
| L | Scanning Electron | 5.9 nm − 19.9 nm | 2 % | 1 % |
| W | Microscope | 2.27 nm | 2 % | 1 % |
| $k_B T$ | Typical value @ 298 K | $4.11 \cdot 10^{-21}$ J | 2 % | 1 % |
| ρ | Cleanroom database | 21450 / 9680 kg/m$^3$ | - | - |
| t | Filmetrics F54 | 20 nm / 52 nm | 0.2 % | 0.1 % |
| Gain | Amplifier specifications | 998 | 0.1 % | 0.1 % |
| $V_{motional}$ | Motional signal | 0.5 mV | 4%-8% | 4%-8% |
| $C_{device}$ | | ~20 pF | | |
| $C_{connector}$ | Measured with LCR meter | 1.36 pF | 0.3 % | 0.3 % |
| $C_{ampli}$ | | 8 pF | | |
| Total Contribution | | | | 8.6% - 12.6% |

## *1.6.3 Flexoelectric tensor contributions to the effective flexoelectric coefficient*

For pure bending of a beam the flexoelectric polarization, in response to the gradient of axial normal strain $\varepsilon_{11}$ in the thickness direction, can be simplified as:

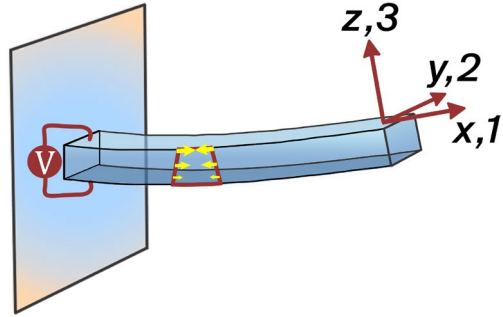

$$P_3 = \mu_{11}\frac{\partial \varepsilon_{33}}{\partial z} + \mu_{12}\left(\frac{\partial \varepsilon_{11}}{\partial z} + \frac{\partial \varepsilon_{22}}{\partial z}\right) \qquad (1.16)$$

$$\begin{cases} \varepsilon_{11} = \dfrac{\sigma_{11}}{E} - \dfrac{\nu\sigma_{22}}{E} - \dfrac{\nu\sigma_{33}}{E} \\ \varepsilon_{22} = -\dfrac{\nu\sigma_{11}}{E} + \dfrac{\sigma_{22}}{E} - \dfrac{\nu\sigma_{33}}{E} \\ \varepsilon_{33} = -\dfrac{\nu\sigma_{11}}{E} - \dfrac{\nu\sigma_{22}}{E} + \dfrac{\sigma_{33}}{E} \end{cases} \qquad (1.17)$$

| **Cantilevers are narrow** $t \ll W \ll L$ | **Cantilevers are wide** $t \ll L \lesssim W$ |
|---|---|
| $\sigma_{22} = 0;\ \sigma_{33} = 0$ | $\varepsilon_{22} = 0;\ \sigma_{33} = 0$ |
| $\begin{cases} \varepsilon_{11} = \dfrac{\sigma_{11}}{E} \\ \varepsilon_{22} = -\dfrac{\nu\sigma_{11}}{E} = -\nu\varepsilon_{11} \\ \varepsilon_{33} = -\dfrac{\nu\sigma_{11}}{E} = -\nu\varepsilon_{11} \end{cases}$ | $\begin{cases} \varepsilon_{11} = \dfrac{\sigma_{11}}{E} - \dfrac{\nu^2\,\sigma_{11}}{E} = \dfrac{1-\nu^2}{E}\sigma_{11} \\ \varepsilon_{22} = 0 \to \sigma_{22} = \nu\sigma_{11} \\ \varepsilon_{33} = -\dfrac{(1+\nu)\nu}{E}\sigma_{11} = -\dfrac{\nu}{1-\nu}\varepsilon_{11} \end{cases}$ |
| $P_3 = \mu_{11} \cdot \dfrac{\partial(-\nu\varepsilon_{11})}{\partial z} + \mu_{12}\left(\dfrac{\partial \varepsilon_{11}}{\partial z} + \dfrac{\partial(-\nu\varepsilon_{11})}{\partial z}\right)$ | $P_3 = -\mu_{11} \cdot \dfrac{\nu}{1-\nu}\dfrac{\partial \varepsilon_{11}}{\partial z} + \mu_{12}\left(\dfrac{\partial \varepsilon_{11}}{\partial z}\right)$ |
| $P_3 = [-\nu \cdot \mu_{11} + \mu_{12}(1-\nu)]\dfrac{\partial \varepsilon_{11}}{\partial z}$ | $P_3 = \left[-\mu_{11} \cdot \dfrac{\nu}{1-\nu} + \mu_{12}\right]\dfrac{\partial \varepsilon_{11}}{\partial z}$ |
| $\mu_{\text{eff,n}} = -\nu \cdot \mu_{11} + \mu_{12}(1-\nu)$ | $\mu_{\text{eff,w}} = \mu_{11} \cdot \dfrac{\nu}{\nu - 1} + \mu_{12}$ |

### 1.6.4 Equivalent flexoelectric charges for the measured voltages

Here, just for curiosity, we calculate what are the equivalent flexoelectric charges for the measured flexoelectric voltages $V_{motional}$.

$$I_{canti} = j\omega \cdot C_{load} \cdot V_{motional} = \frac{dQ}{dt} = j\omega Q \quad (1.18)$$

$$Q = C_{load} \cdot V_{motional} \quad (1.19)$$

The measured $C_{load} = 29.09\ pF$.

The equivalent charge for the flexoelectric measurements we show in the paper (cantilever of 40 μm width). Charge of an electron $1.602 \cdot 10^{-19}\ C$.

| $V_{motional}$ [μV] | Q [C] | **Equivalent** electron charge at resonance frequency |
|---|---|---|
| 0.85 | $2.5 \cdot 10^{-17}$ | 154 e⁻ |
| 0.56 | $1.6 \cdot 10^{-17}$ | 101 e⁻ |
| 0.28 | $8.1 \cdot 10^{-18}$ | 50 e⁻ |
| 0.14 | $4.1 \cdot 10^{-18}$ | 25 e⁻ |

For the case of the 10 μm width cantilever, the $C_{load} = 31.9$ pF.

| $V_{motional}$ [μV] | Q [C] | **Equivalent** electron charge at resonance frequency |
|---|---|---|
| 0.202 | $6.4 \cdot 10^{-18}$ | 40 e⁻ |
| 0.169 | $5.4 \cdot 10^{-18}$ | 34 e⁻ |
| 0.143 | $4.6 \cdot 10^{-18}$ | 28 e⁻ |
| 0.102 | $3.3 \cdot 10^{-18}$ | 20 e⁻ |
| 0.069 | $2.2 \cdot 10^{-18}$ | 14 e⁻ |
| 0.034 | $1.1 \cdot 10^{-18}$ | 7 e⁻ |

Our methodology does not measure flexoelectric charges directly, we measure currents, which is an integration of the charges at a relatively high frequency (200 kHz). We found these calculations interesting to have an estimated electron charge produced by the flexoelectric effect for hafnium oxide.